\documentclass[prl,showpacs,twocolumn]{revtex4}
\usepackage{graphics}
\usepackage{epsfig}
\usepackage{color}

\begin{document}
\title{The Repressor-Lattice:\\ Feedback, Commensurability, and Dynamical Frustration.}
\author{Mogens H.~Jensen, Sandeep Krishna and Simone Pigolotti}

\email{mhjensen@nbi.dk}
\affiliation{Niels Bohr Institute, Blegdamsvej 17, DK-2100,
Copenhagen, Denmark} \homepage{http://cmol.nbi.dk}
\date{\today}

\begin{abstract}
  We construct a hexagonal lattice of repressing genes, such that each
  node represses three of the neighbors, and use it as a model for
  genetic regulation in spatially extended systems.  Using symmetry
  arguments and stability analysis we argue that the repressor-lattice
  can be in a non-frustrated oscillating state with only three
  distinct phases. If the system size is not commensurate with three,
  oscillating solutions of several different phases are possible. As
  the strength of the interactions between the nodes increases, the
  system undergoes many transitions, breaking several
  symmetries. Eventually dynamical frustrated states appear, where the
  temporal evolution is chaotic, even though there are no built-in
  frustrations. Applications of the repressor-lattice to real
  biological systems are discussed.
\end{abstract}


\maketitle

Our understanding of genetic regulation inside the cell has greatly
improved in recent years. A number of genetic circuits have been
quantitatively characterized, ranging from switches to oscillators
made up of negative feedback loops. The latter class of circuits is
ubiquitous in regulatory networks with oscillating gene expressions,
two of the most important examples being the NFkB network for
inflammatory response \cite{Nelsonetal,Hoffmann,sandeep_pnas} and the
p53-mdm2 system which regulates cell apoptosis \cite{p53,tiana}.

However, decisions taken inside the cell may depend crucially on the
environment and may be cooperative, i.e. depend on the behavior of
neighboring cells. This calls for theoretical modeling which
explicitly takes the spatial arrangement of different cells into
account.  As a basic unit, we consider a negative feedback loop
consisting of three proteins that repress each other by blocking the
associated genes, which Leibler and Elowitz termed the 'repressilator'
\cite{EL}.  Previously, others have studied coupled repressilators to
investigate quorum sensing \cite{strogatz} and
cell-to-cell communication \cite{population}. As a further step, one
might consider systems made up of regular arrays of cells interacting
in a specific manner with neighboring cells. Because of close packing,
real arrays of cells in planar tissues often display hexagonal or
near-hexagonal structure, e.g. in hepatic or retinal tissue
\cite{hex1,hex2,hex3}.

\begin{figure}[htbp]
\center\epsfig{file=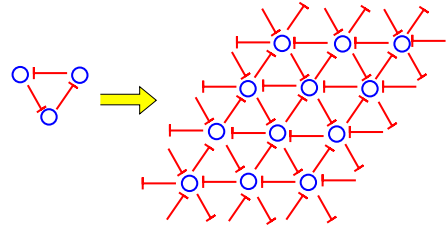,angle=0,width=5.8cm}
\caption{The construction of the repressor-lattice from 'units' of
single repressilators suitably placed on a hexagonal lattice. Each
link symbolizes a repressor between two nodes corresponding to
repressing genes, proteins, species, etc.} 
\label{lattice}
\end{figure}

Here we approach this general problem by extending the simple
repressilator to a repressor-lattice -- a hexagonal array of repeated
and overlapping repressilator motifs, as shown in Fig. \ref{lattice}.
Each node is repressed by three neighboring nodes and at the same time
represses three other neighbors. A biological implementation of such a
system would require a tissue where cells communicate specifically
with their immediate neighbours, rather than in a mean-field manner as
in quorum sensing. Such direct communication is in fact quite common,
either through small conduits that connect the cells, or via proteins
that span the membrane of the cells \cite{juxtacrine}. Further, the
directed nature of the interactions would require some form of
epigenetic gene silencing, resulting in adjacent cells expressing
different genes even though they have exactly the same DNA
\cite{birdwolffe,suzuki,mille}.  The modeling framework we propose is,
however, general and can be used to describe other kind of interations,
such as bidirectional ones, which might be easier to realize
experimentally.

The lattice in Fig. \ref{lattice} can be naturally
constructed to be translationally invariant and such that all local
loops are repressilator motifs. We will approach the problem by
imposing periodic boundary conditions in order to preserve
translational invariance.  Later, we will discuss how these results
translate to the case of a large lattice without the
periodic boundary conditions, which is more relevant for biology. 

The basic repressilator motif may exhibit an oscillating state with a
phase difference between consecutive variables equal to 2$\pi$/3. One
can ask whether the entire repressor-lattice might exist in an
oscillatory state where only three different phases are allowed, each
differing by 2$\pi$/3.  We will show that this is indeed the case, but
lattice commensurability effects may break this scenario.

\begin{figure}[tbp]
{
\epsfig{width=.32\textwidth,file=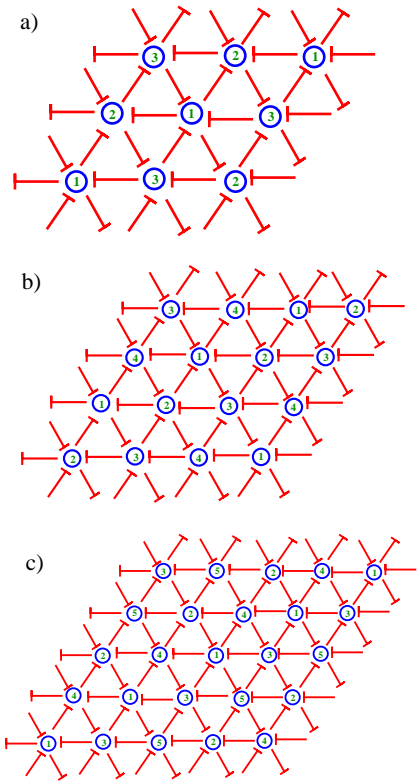,width=5.5cm,angle=0}
}
\caption{Systems of 3 $\times$ 3 (a), 4 $\times$ 4 (b) 
and 5 $\times$ 5 (c) nodes subjected to
periodic boundary conditions as indicated by the extra links.
The numbers refer to the different phases of the solutions just above
Hopf bifurcations. In (a) the solution exhibits symmetry with respect
to rotations of angles which are multiples of $2\pi/3$. In (b), (c) 
this symmetry is broken, so that $3$ distinct solutions coexist above 
the Hopf bifurcation. }
\label{unit}
\end{figure}

In the repressor-lattice, the variable at a node $(m,n)$ is repressed
by three neighboring nodes which we represented by an interaction term
$F_{int}$, leading to a dynamical equation for the concentration
of species $x_{m,n}$:
\begin{equation}
\frac{dx_{m,n}}{dt} = c - \gamma ~x_{m,n} + \alpha F_{int}
\label{rep}
\end{equation}
We consider two types of interaction terms -- either an additive
repression (an 'or gate'):
\begin{equation}
F_{int} = \frac{1}{1 + (\frac{x_{m+1,n}}{K})^h} 
+ \frac{1}{1 + (\frac{x_{m,n-1}}{K})^h} +
\frac{1}{1 + (\frac{x_{m-1,n+1}}{K})^h}
\label{add}
\end{equation}
or a multiplicative repression (an 'and gate'):
\begin{equation}
F_{int} = \frac{1}{1 + (\frac{x_{m+1,y}}{K})^h} 
\cdot \frac{1}{1 + (\frac{x_{m,n-1}}{K})^h}
\cdot \frac{1}{1 + (\frac{x_{m-1,n+1}}{K})^h}~~.
\label{mult}
\end{equation}

In either case we use standard Michaelis-Menten terms to model the
repression.  The parameter $c$ measures the constitutive production of
the proteins, $\gamma$ determines the degradation rate and $\alpha$
the strength of the repression by another protein.  Further, $K$ is
the dissociation constant of the binding complex whereas $h$ is the
Hill coefficient measuring its cooperativity.
For simplicity we assign the same parameter values to all the nodes in
the lattice.  We note that Ref. \cite{EL} also introduced the
associated mRNA for each gene resulting in six coupled ordinary
differential equations. For simplicity we keep only the protein
variables leading to three coupled equations - a single repressilator
with this simplification can still be brought into an oscillating
state \cite{simone}.

\begin{figure}[htbp]
{
\epsfig{width=.32\textwidth,file=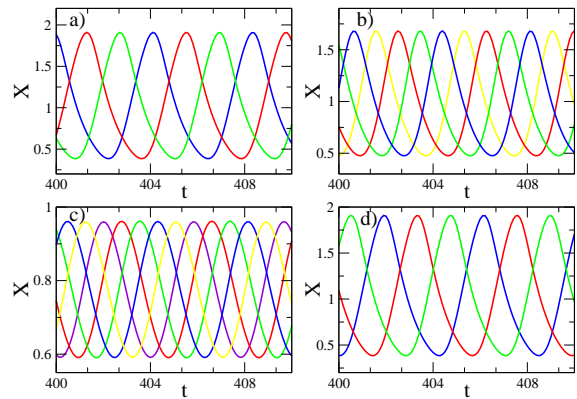,width=7.5cm,angle=0}
}
\caption{Solutions of repressor-lattices of sizes
a): 3 $\times$ 3, b): 4 $\times$ 4, c): 5 $\times$ 5, d): 6 $\times$ 6
with multiplicative interactions, Eq. (\ref{mult}) and parameters
  $c=0.1, \gamma=1.0, K=1.0, h=2$. The value of $\alpha$ is in
each case chosen to be just above the Hopf bifurcation.
Note that three, four and five different phases exist 
in a), b), c), respectively.
In d) there are however only three different phases.}
\label{multfig}
\end{figure}

For a single repressilator there exists a large regime of parameter
space where oscillations are possible
\cite{EL,simone}. The transition to oscillations occurs via a Hopf
bifurcation.  We find similar behavior in the repressor-lattice.  As
a starting point, a lattice with 3 $\times$ 3 nodes as in
Fig. \ref{unit}a was simulated both with additive, Eq. (\ref{add}),
and multiplicative, Eq. (\ref{mult}), couplings.  Just above the Hopf
bifurcations, we found smooth oscillations with only three distinct
phases as indicated by the numbers 1,2,3 labeling the nodes in
Fig. \ref{unit}a. The oscillating time series is shown in
Fig. \ref{multfig}a.  These solutions are trivially related to the
solutions of the basic repressilator motif since each node receives
three identical inputs, with a $2\pi/3$ phase shift with respect to
itself. Note that this solution is invariant under lattice rotation of
multiples of $2\pi/3$

However, this scenario is not completely general.  For instance, in
the case of a lattice of size 4 $\times$ 4 (16 nodes) the
corresponding dynamical solutions are different, as shown in
Fig. \ref{multfig}b. As in the previous case (Fig. \ref{multfig}a) we
are relatively close to the first Hopf bifurcation. However, now
phases of the oscillating solutions differ by 2$\pi$/4 between the nodes.  
The origin of this is a
commensurability effect between the number of nodes in the lattice and
the associated number of possible phases of the oscillating
solutions. This commensurability effect is of course enforced by the
periodic boundary conditions. The complete structure of the phases is
shown on the 4 $\times$ 4 unit cell in Fig. \ref{unit}b. The
case of 5 $\times$ 5 is also affected by commensurability effects,
as shown in Figs. \ref{unit}c and \ref{multfig}c.

As opposed to the 3 $\times$ 3 system, here the inputs arriving to a
specific node are different. This reflects the fact that the
oscillatory solution is no longer rotationally invariant.  We note
that all lattices which are commensurate by three, i.e. 6 $\times$ 6
(see Fig. \ref{multfig}d),
9 $\times$ 9, etc, allow a non-frustrated, symmetric state similar to
the 3 $\times$ 3 system.  These periodic solutions all exhibit a
Goldstone mode in the sense that it is possible to slide the phases as
long as the phase differences are kept constant. This means that the
specific values of the phases for the solutions are determined by the
initial conditions.

In order to understand these solutions in depth, we perform a
stability analysis. We consider the ``or" gate
Eqs. \ref{rep},\ref{add} and, since the system is translationally
invariant, we search for a constant homogeneous solution:
\begin{equation}\label{homogeneous}
x_{m,n}=x^*~~ \forall m, n\ \longrightarrow\  3 \alpha K^h=(\gamma x^*-c) 
(K^h+{x^*}^h).
\end{equation}
The equation for $x^*$ always has one, and only one, real positive
solution. The next step is to perturb the homogeneous solution in order to
perform a stability analysis. We consider a first order perturbation of the
form:
\begin{equation}
x_{m,n}(t)=x^*+\epsilon \exp\left[\lambda t + \frac{2\pi i(k_m m+k_n n)}{L}
\right]~~.
\end{equation}
Notice that since the solution must have the periodicity of the
lattice, $k_m$ and $k_n$ should be natural numbers and also $1\le
k_m,k_n\le L$. Plugging the solution into Eq. (\ref{rep}) and
expanding to first order in $\epsilon$ yields the following
dispersion relation:
\begin{equation}\label{dispersion}
\lambda=-\tilde a \left(e^\frac{2\pi i k_m}{L}+ e^{-\frac{2\pi i k_n}{L}}+e^\frac{2\pi i (k_n-k_m)}{L}\right)-\gamma,
\end{equation}
where $\tilde{a}=\alpha hK^h(x^*)^{h-1}/[K^h+(x^*)^h]^2 $. Other kind
of interaction terms lead to the same dispersion relation simply with
a slightly different definition of $x^*$ and $\tilde{a}$; for example,
taking multiplicative interactions leads to $\tilde{a}=\alpha
(h/K)(x^*/K)^{h-1}/[1+(x^*/K)^h]^4$. Eigenvalues $\lambda$ with a
positive real part will destabilize the homogeneous solution. Taking
the real part of expression (\ref{dispersion}), the eigenvalue with
the largest real part is the one that minimizes the function:
\begin{eqnarray}
f(k_m,k_n) =\qquad\qquad\qquad\qquad\qquad\qquad\qquad\qquad\qquad\nonumber\\
\cos\left(\frac{2\pi k_n}{L}\right)+ 
\cos\left(-\frac{2\pi k_m}{L}\right)+ 
\cos\left(\frac{2\pi (k_m-k_n)}{L}\right).
\label{min}
\end{eqnarray}
Before finding the solutions, we stress that
$f(k_m,k_n)=f(-k_m,-k_n)$, while the imaginary part of the eigenvalue
changes sign when the wave vector changes sign. This means that the two
vectors $(k_m,k_n)$ and $(-k_m,-k_n)$ minimizing the function $f$ are
the complex conjugate pair that will cause the Hopf
bifurcation. The function $f$ is independent of the
parameters of the system, meaning that the kind of pattern 
depends not on the form of the interaction (as long as the lattice is
homogeneous with the same geometry), but on the number of sites
in the lattice. The value of $\gamma$ determines only
how much we have to increase $\tilde{a}$ to encounter the Hopf
bifurcation.

\begin{figure}[htbp]
\center\epsfig{file=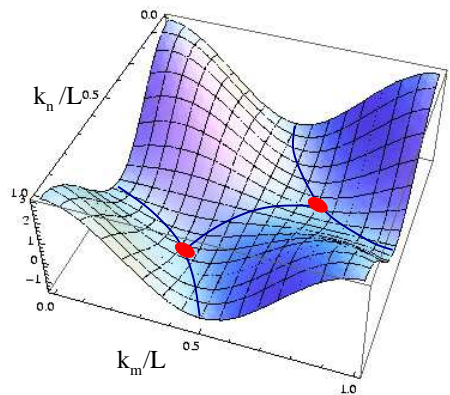,angle=0,width=6cm}
\caption{The landscape of the function Eq. (\ref{min}). Red dots mark 
absolute minima, corresponding to the symmetric solution for $L$ 
multiple of $3$. Blue lines mark the bottom of the valleys around 
the minima. When $L$ is not multiple of $3$, the absolute minima are 
not achievable, and the degenerate solutions are given by $3$ complex 
conjugate pairs along these valleys.}
\label{landscape}
\end{figure}

Plotting the function $f(k_m/L,k_n/L)$ in the first Brillouin zone,
$0 \le k_m,k_n<L$ we see that it achieves its absolute minimum for the
couple of eigenvalues $(k_m/L,k_n/L)=(1/3,2/3)$ and
$(k_m/L,k_n/L)=(2/3,1/3)$, where $f(k_m,k_n)=-3/2$, see
Fig. \ref{landscape}. This means that a
Hopf bifurcation will occur when $3\tilde{a}=2\gamma$. Of course these
wave vectors are allowed only when $L$ is a multiple of $3$, so that
the values of $k_m$ and $k_n$ at the minimum are natural numbers.

We can of course minimize the function $f$ also for values of $L$ that are not
multiples of $3$. For $L=4$, the minimum is $f(1,3)=f(3,1)=-1$, but
also $f(2,3)=f(3,2)=-1$ and $f(3,4)=f(4,3)=-1$. The case $L=5$ is also
a degenerate case. The minimum is $f(2,3)=f(3,2)\approx -1.30902$, but
also $f(1,3)=f(4,2)\approx -1.30902$ and $f(2,4)=f(3,1)\approx
-1.30902$. All cases that are not multiples of $3$ have this
degeneracy, due to the symmetry of the lattice. Close to the Hopf
bifurcation, the number of observed phases will reflect the
periodicity of the eigenfunction. In particular, there will always be
$3$ distinct phases if $L$ is a multiple of $3$ and $L$ phases if $L$ is
a prime number.

The phases of the eigenfunctions can be used to figure out how the
oscillation pattern will look like on the lattice: sites on the
lattice at a distance $\Delta m, \Delta n$ such that $k_m \Delta m +
k_n \Delta n = 0$ will be in phase. Fig. \ref{unit}c shows one of
these solutions of the 5 $\times$ 5 lattice, namely $f(4,1)$.
The other 'symmetric' solutions can be obtained through rotations of
multiples of $2\pi/3$, respecting the hexagonal rotational symmetry of
the lattice.
\begin{figure}[htbp]
\vbox{
{
\epsfig{file=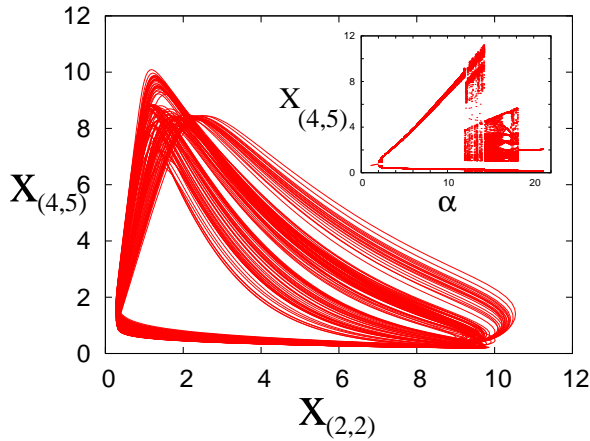,width=0.45\textwidth,angle=0}
}
}
\caption{
A specific chaotic solution for a 5 $\times$ 5 lattice
with multiplicative coupling
(Lyapunov exponent equal to 0.028)
obtained at a
coupling strength equal to $\alpha=12.8$ with the variable in node $(4,5)$ 
plotted against the variable in node $(2,2)$.
Other parameter values are: $c=0.1,\gamma=1.0,K=1.0,h=3$.
Inset: Bifurcation diagram, showing 
the maxima and minima for dynamical solutions in the node point
$(4,5)$ after a transient period of 15000 time units. The extremal
values are plotted against varying values of the coupling strength $\alpha$.
The critical value for the Hopf bifurcation calculated from the dispersion
relation \ref{dispersion} (see main text) gives $\alpha_c = 1.838$.
}
\label{chaos}
\end{figure}

One might expect that the symmetric solutions with five different
phases, Fig. \ref{unit}c, could exist even when parameters are varied.
This is not the case: when the coupling parameter $\alpha$ is
increased, several transitions related to strong non-linear effects
take place.  Starting out with smooth periodic solutions of five
distinct phases, amplitude modulations set in when the coupling
constant $\alpha \approx 2.6$.  At the same parameter value, we also
observe that some phases that were distinct before this transition now
coalesce with each other.  At higher $\alpha$-values, amplitude
modulations become even more pronounced and furthermore temporal
period-doubling sets in. Increasing $\alpha$ even more, chaotic
solutions eventually appear as shown in the bifurcation diagram and
the attractor of Fig. \ref{chaos} (similar bifurcation diagrams
are observed for repressive cell-cell communication \cite{population}). 
Even though the lattice is made up
of simple repressilators without local frustration, the resulting
dynamics is chaotic: it is not possible to keep the simple five-phase
solutions when the repressilators are coupled strongly 
with the neighbors.  Each node in the 5 $\times$ 5 lattice exhibits a
different bifurcation diagram (no simple period five symmetry
operations are present) showing that all symmetries are eventually
broken through a series of non-linear transitions.


One may wonder how realistic periodic boundary conditions are for
modeling real biological systems. In the $3\times3$ case, this might
be implemented in a single cell with $9$ different genes, each
repressed by three different ones. Having in mind extended systems, a
more realistic case is to consider a large, finite lattice with
non-periodic boundary conditions to represent an isolated planar
tissue, in which cells at the boundary receive no external signal.  We
found from simulations that such a system shows frustration effects
similar to the case of periodic boundaries: when the steady state is
destabilized, cells far from the boundaries exhibit the three-phase
dynamics of the repressilator circuit, while closer to the boundaries
the dynamics is more irregular, with more phases possible. We did not
observe any chaos in this case, even for very large values of
$\alpha$.

In conclusion, the lattice model we have investigated here provides a
simple starting point to study regulation in spatially extended
biological systems.  Future direction could include, for instance,
introducing an intrinsic 'frustration' in the repressor-lattice.
There are several ways of doing this, e.g. by lattice defects, or
mutations modifying some of the interactions.  For example, one can
consider what happens when a specific repressor link is mutated into
an activator.  These generalizations may provide a useful framework
for describing more specific cases of cell-to-cell communication in
biological tissues.

We are grateful to Namiko Mitarai, Joachim Mathiesen and Szabolcs
Semsey for discussions. This work
is supported by Danish National Research Foundation.

\end{document}